%% file: main.tex
\documentclass[conference]{IEEEtran}
\IEEEoverridecommandlockouts

\usepackage{cite}
\usepackage{amsmath,amssymb,amsfonts}
\usepackage{graphicx}
\usepackage{textcomp}
\usepackage{xcolor}
\usepackage{booktabs}
\usepackage{refcount}
\usepackage[caption=false,font=footnotesize]{subfig}
\usepackage{algorithm}
\usepackage{algpseudocode}
\algrenewcommand\algorithmicrequire{\textbf{Input:}}
\algrenewcommand\algorithmicensure{\textbf{Output:}}
\usepackage[hidelinks]{hyperref}
\usepackage{eso-pic}
\newtheorem{lemma}{Lemma}

\graphicspath{{figures/}}
\newcommand{\figeqcheck}[2]{\IfRefUndefinedExpandable{#1}{}{\ifnum\getrefnumber{#1}=#2\relax\else
  \leavevmode{\color{gray}\footnotesize[Fig.~\ref{fig:receiver} shows (#2) for #1, now (\getrefnumber{#1}): recompile figures/F1\_receiver.tex]}\par\fi}}

\newif\ifshownotes
\shownotestrue

\newcommand{\T}{{\sf T}}
\newcommand{\Hr}{{\sf H}}
\newcommand{\bH}{\mathbf{H}}
\newcommand{\bY}{\mathbf{Y}}
\newcommand{\bX}{\mathbf{X}}
\newcommand{\bW}{\mathbf{W}}
\newcommand{\bG}{\mathbf{G}}
\newcommand{\bI}{\mathbf{I}}
\newcommand{\bh}{\mathbf{h}}
\newcommand{\bb}{\mathbf{b}}
\newcommand{\bq}{\mathbf{q}}
\newcommand{\bJ}{\mathbf{J}}
\newcommand{\bx}{\mathbf{x}}
\newcommand{\by}{\mathbf{y}}
\newcommand{\bw}{\mathbf{w}}
\newcommand{\bv}{\mathbf{v}}
\newcommand{\br}{\mathbf{r}}
\newcommand{\bu}{\mathbf{u}}
\newcommand{\bA}{\mathbf{A}}
\newcommand{\bC}{\mathbf{C}}
\newcommand{\bD}{\mathbf{D}}
\newcommand{\bU}{\mathbf{U}}
\newcommand{\vm}{\mathbf{m}}
\newcommand{\bLambda}{\boldsymbol{\Lambda}}
\newcommand{\bSigma}{\boldsymbol{\Sigma}}
\newcommand{\bfeta}{\boldsymbol{\eta}}

\newcommand{\bXi}{\boldsymbol{\Xi}}

\newcommand{\ours}{proposed prior}
\newcommand{\gmm}{GMM prior}
\newcommand{\gauss}{Gaussian prior}
\newcommand{\pcsi}{Perfect CSI}
\newcommand{\unrep}{unrepaired diffusion prior}
\newcommand{\scdiff}{scalar-site diffusion prior}

\begin{document}
\AddToShipoutPictureFG*{\AtPageUpperLeft{\put(0,-0.42in){\makebox[\paperwidth]{\parbox{0.8\paperwidth}{\centering\footnotesize This work has been submitted to the IEEE for possible publication. Copyright may be transferred without notice, after which this version may no longer be accessible.}}}}}
\bstctlcite{IEEEexample:BSTcontrol}

\title{Learned Diffusion Channel Priors for Coded MIMO Turbo Receivers
\thanks{This work was supported by Institute of Information \& communications Technology Planning \& Evaluation (IITP) under 6G Cloud Research and Education Open Hub (IITP-2025-RS-2024-00428780) grant funded by the Korea government (MSIT).}
}

\author{
\IEEEauthorblockN{Taejun Ha\IEEEauthorrefmark{1}, Hyeji Kim\IEEEauthorrefmark{2}, and Jeonghun Park\IEEEauthorrefmark{1}}
\IEEEauthorblockA{\IEEEauthorrefmark{1}\textit{Yonsei University}, \IEEEauthorrefmark{2}\textit{The University of Texas at Austin}\\
Emails: tjha@yonsei.ac.kr, hyeji.kim@austin.utexas.edu, jhpark@yonsei.ac.kr}
}

\maketitle

\begin{abstract}
Iterative receivers for coded multiple-input multiple-output (MIMO) links estimate the channel from pilots, decoded data, and a prior.
In short packets with few pilots and at low signal-to-noise ratio (SNR), the prior matters.
An expectation propagation (EP) turbo receiver needs the channel's mean and covariance given an anisotropic Gaussian message from the pilots and decoder.
A diffusion prior's denoiser, however, returns only a mean, and only for a channel in isotropic noise.
We replace only the receiver's channel-prior module with a learned diffusion prior.
At an isotropic query formed from the message, one network gives both moments: its output is the mean, its Jacobian the covariance.
They form a matrix-valued EP site, the Gaussian factor representing the prior.
We show why an indefinite Jacobian can make the receiver diverge, and we floor its eigenvalues.
In coded block error rate (BLER), the diffusion prior fails less than a Gaussian mixture model (GMM) prior learned from the same channels.
It does so on a controlled sparse specular model and a 3GPP model, at every tested receive-array and training-set size.
On the former, scalar-site controls, with one variance instead of the covariance, keep this advantage, so the gain comes from the diffusion prior.
\end{abstract}

\begin{IEEEkeywords}
diffusion, expectation propagation, Gaussian mixture models, MIMO channel estimation, turbo receivers
\end{IEEEkeywords}

\section{Introduction}\label{sec:intro}

\begin{figure}[t]
\centering
\includegraphics[width=0.9\columnwidth]{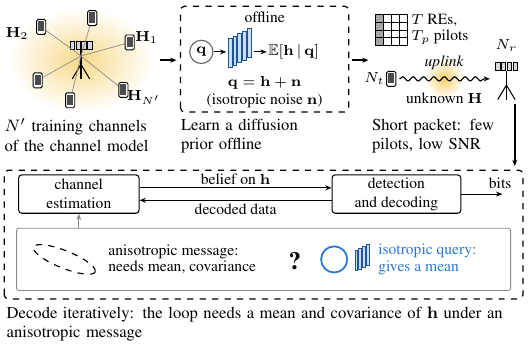}
\caption{Problem setting: a prior learned offline from $N'$ training channels must serve the iterative receiver of a short packet (Section~\ref{sec:problem}).}
\label{fig:problem}
\end{figure}

An iterative receiver for a coded multiple-input multiple-output (MIMO) link estimates the channel from the pilots, the decoded data, and a prior on the channel, here learned offline from training channels (Fig.~\ref{fig:problem}).
With few pilots and at low signal-to-noise ratio (SNR), much of the channel stays uncertain, so the prior matters.
Short packets afford few pilots: a packet spans few resource elements~\cite{durisi2016toward}, and every pilot takes one of them from the data~\cite{ostman2019short}.
Decoded data of the same packet can refine the pilot estimate~\cite{valenti2001iterative}, but at low SNR they are uncertain too.

In a turbo receiver based on expectation propagation (EP), the channel-prior module must compute the channel's posterior mean and covariance given a Gaussian message from the pilots and the decoder (Fig.~\ref{fig:problem}).
This message amounts to a weighted least-squares channel estimate and its error covariance.
As the receiver iterates between channel estimation, detection, and decoding, the decoded data make this message anisotropic: its covariance is no longer a multiple of the identity.
EP represents the prior by a Gaussian factor called a site, and a matrix-valued site carries a full covariance rather than a scalar variance.

Learned priors have been used to obtain a point estimate without a covariance and, in message passing, a scalar variance in place of the covariance.
A diffusion prior is represented by a denoising network, which returns only a mean, and only for a channel observed in isotropic noise of a given variance.
Score-based and diffusion estimators recover the channel from the pilots~\cite{arvinte2023mimo,fesl2024diffusion} or jointly with data detection~\cite{zilberstein2024joint}.
One receiver feeds soft data decisions back to a conditional diffusion estimator, with no decoder in the loop, and its detector uses a point channel estimate~\cite{zhou2026mimo}.
In message-passing algorithms~\mbox{\cite{cai2025joint,cai2025score,wadayama2026score}}, learned denoisers pass a scalar variance.
We are not aware of a coded receiver that estimates its channel with a diffusion denoiser used as a matrix-valued EP site.

No diffusion prior has been compared, to our knowledge, in coded block error rate (BLER) with its natural baseline, a Gaussian mixture model (GMM) prior learned from the same training channels.
This baseline gives the required mean and covariance exactly, from the anisotropic message itself.
GMM priors give closed-form channel estimators, including data-aided ones~\cite{koller2022asymptotically,weisser2023data}, and have been compared with a diffusion prior, but only for pilot-based estimation~\cite{fesl2024diffusion}.

In this paper, we replace the channel-prior module of an EP turbo receiver (module~H in Fig.~\ref{fig:receiver}) with a learned diffusion prior.
A diffusion prior, unlike a GMM prior, does not restrict the channel distribution to a finite Gaussian mixture.
Module~H condenses the anisotropic message into an isotropic query, one observation of the channel in isotropic noise, and evaluates the network there (Section~\ref{sec:site}).
The noise variance of the query changes with the SNR and the iteration.
The network learns to denoise the channel at a continuum of noise levels, so by Tweedie's formula~\cite{efron2011tweedie} its output approximates the posterior mean given the query.
One network is thus used at every iteration and SNR, whereas~\cite{arvinte2023mimo} draws posterior samples.
For an exact denoiser, the Jacobian of its output with respect to the query, scaled by the noise variance, is the posterior covariance, as in~\cite{rozet2024learning}.
We turn the mean and covariance into a matrix-valued EP site, so that the full covariance reaches the detector.

Our contributions are as follows:
\begin{itemize}
\item We turn a learned diffusion denoiser into a matrix-valued EP site with the same input and output as the closed-form site of the \gmm{} (Section~\ref{sec:site}).
The Jacobian of a learned network can be indefinite.
We show why this can make the receiver diverge even when the eigenvalues of the site precision, the site's inverse covariance, are floored.
We instead floor the eigenvalues of the Jacobian, which bounds the site (Section~\ref{sec:psd}).
\item We compare the \ours{} in coded BLER with a \gmm{} learned from the same training channels, using paired sign tests on the same test blocks (Section~\ref{sec:setup}).
On a controlled sparse specular model (Sparse specular) with 8 receive and 4 transmit antennas ($8\times4$), it fails less at every tested training-set size.
At the default size, it needs 1.41\,dB less SNR for BLER 0.1 (Section~\ref{sec:headline}).
It also fails less than each of 16 baselines, including three sparse estimators.
We further test the receiver with 32 receive antennas on 3GPP UMi 28\,GHz, the urban micro (UMi) model of the 3rd Generation Partnership Project (3GPP).
There, the \ours{} still fails less than a \gmm{} with up to 8192 components and needs 2.17\,dB less SNR for BLER 0.1.
\item We identify where the gain comes from (Section~\ref{sec:source}).
In a control on Sparse specular $8\times4$, we give the diffusion prior only a scalar site, one variance in place of the covariance.
It still fails less than the \gmm{} with its matrix site, so the gain over the \gmm{} comes from the learned diffusion prior in module~H.
On Sparse specular and 3GPP UMi 28\,GHz, both at $8\times4$, the \ours{} fails less in the turbo loop than when it is applied once to the pilots.
\end{itemize}

\section{System Model and Problem Statement}\label{sec:system}

\subsection{Signal Model}\label{sec:signal}
We consider a coded MIMO link in which an $N_t$-antenna transmitter sends one short packet per block of $T$ resource elements (REs) to an $N_r$-antenna receiver (an $N_r\times N_t$ link).
We model a block as one coherence region of an orthogonal frequency-division multiplexing (OFDM) grid, e.g., 4 subcarriers by 4 OFDM symbols.
The channel $\bH\in\mathbb{C}^{N_r\times N_t}$ is constant within a block and independent across blocks, and the received block is
\begin{align}
\bY&=\bH\bX+\bW,\qquad \bX=[\bX_p\;\bX_d]\in\mathbb{C}^{N_t\times T},\label{eq:signal}\\
\by_t&=(\bx_t^{\T}\otimes\bI_{N_r})\,\bh+\bw_t,\label{eq:column}
\end{align}
where $\bX_p$ holds $T_p$ pilot columns, $\bX_d$ holds $T_d=T-T_p$ data columns, $\bW$ has independent and identically distributed (i.i.d.) $\mathcal{CN}(0,\sigma^2)$ entries, and \eqref{eq:column} is column~$t$ of \eqref{eq:signal} with $\bh=\mathrm{vec}(\bH)\in\mathbb{C}^{N}$ and $N=N_rN_t$.

Each block carries $T_p$ pilot columns and one codeword, so the BLER is a per-codeword rate.
The $T_p=N_t$ pilot columns form the $N_t$-point discrete Fourier transform (DFT) matrix, so $\bX_p\bX_p^{\Hr}=T_p\bI$.
The data carry $K_{\mathrm b}=N_tT_d-6$ information bits, encoded by the 64-state rate-1/2 $(133,171)_8$ convolutional code with six zero tail bits.
The code bits are mapped to quadrature phase-shift keying (QPSK) symbols and permuted by a per-block random symbol interleaver~$\pi$.
Data symbols have unit energy, every RE carries total energy $N_t$, and $\mathbb{E}\|\bH\|_F^2=N_rN_t$ for every channel model, so $\mathrm{SNR}=1/\sigma^2$ is the mean SNR of one transmit antenna at one receive antenna.

The receiver knows $\bX_p$, $\sigma^2$, the code, and $\pi$ but not $\bH$; it must decode each block while estimating $\bh$ from the pilots, the decoded data, and a prior (Fig.~\ref{fig:problem}).
All learned priors are fitted offline to the same $N'$ channels drawn from the channel model (training channels), separately from the test blocks.
By default, $N_t=4$, $T=16$, $T_p=4$ ($K_{\mathrm b}=42$), $N_r=8$, and $N'=1.6\times10^5$.

\subsection{Channel Models}\label{sec:channels}
Our default model, Sparse specular, is a controlled sparse specular model that isolates one property: given its paths, the channel is not Gaussian, whereas a GMM is Gaussian given its component.
Per block, it draws $L\sim\mathcal{U}\{3,\dots,8\}$ paths with i.i.d.\ angles of arrival and departure $\theta_\ell,\phi_\ell\sim\mathcal{U}[-\pi/3,\pi/3]$ and forms
\begin{equation}
\bH=\sqrt{N_rN_t}\sum_{\ell=1}^{L}g_\ell\,\mathbf{a}_{N_r}(\theta_\ell)\,\mathbf{a}_{N_t}^{\Hr}(\phi_\ell),
\label{eq:d2}
\end{equation}
where $\mathbf{a}_M(\theta)=M^{-1/2}[e^{j\pi n\sin\theta}]_{n=0}^{M-1}$, and the gains $g_\ell=\sqrt{p_\ell}\,e^{j\psi_\ell}$ have fixed powers $p_\ell\propto e^{-\ell/2}$, $\sum_\ell p_\ell=1$, and phases $\psi_\ell\sim\mathcal{U}[0,2\pi)$ that absorb the path delays.
Because $|g_\ell|$ is fixed, $\bh$ is not Gaussian given the paths: each entry has the normalized fourth moment $\mathbb{E}|h_n|^4/(\mathbb{E}|h_n|^2)^2=2-\sum_\ell p_\ell^2<2$, whereas a complex Gaussian entry has exactly~2.

Section~\ref{sec:results} also uses 3GPP UMi 28\,GHz, narrowband snapshots of the TR~38.901 urban micro (UMi) channel~\cite{3gpp38901} from Sionna~2.1~\cite{hoydis2022sionna}.
For it, $\bX_p$ is instead $\sqrt{N_t}$ times the eigenvectors of the transmit correlation $\mathbb{E}[\bH^{\Hr}\bH]/N_r$ of the training channels, which keeps $\bX_p\bX_p^{\Hr}=T_p\bI$ and hence the same pilot information as DFT pilots.

\subsection{Iterative Receiver}\label{sec:receiver}
\begin{figure}[t]
\centering
\includegraphics[width=0.8\columnwidth]{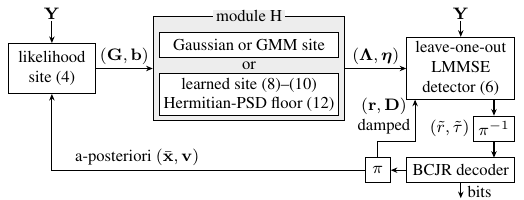}\par
\InputIfFileExists{figures/F1_receiver_eqnums.tex}{}{}%
\caption{Coded turbo receiver (one of 16 outer iterations); the priors differ only in module~H, and $\pi$ is the symbol interleaver.}
\label{fig:receiver}
\end{figure}

To isolate the prior, the receivers with the \gauss{}, \gmm{}, and \ours{} share one turbo loop (Fig.~\ref{fig:receiver}) and differ only in module~H, the channel-prior module.
The loop runs 16 outer (turbo) iterations, each passing once through channel estimation, a linear minimum mean-square error (LMMSE) detector, and a Bahl--Cocke--Jelinek--Raviv (BCJR) decoder~\cite{bahl1974optimal}.

\textit{Channel estimation:}
The receiver's belief on $\bh$, its Gaussian approximation of the posterior, is the product of two Gaussian factors, which EP calls sites~\cite{minka2001expectation}: a likelihood site from the $T$ columns and a prior site from module~H.
Let $(\bar\bx_t,\bv_t)$ be the soft means and variances of the symbols in column $t$: the pilots with $\bv_t=\mathbf 0$, or the decoder's a-posteriori moments (initially $\bar\bx_t=\mathbf 0$ and $\bv_t=\mathbf 1$).
Treating the symbol uncertainty as noise gives the likelihood site $\exp(-\bh^{\Hr}\bG\bh+2\Re\{\bb^{\Hr}\bh\})$ with
\begin{align}
\bG&=\textstyle\sum_{t=1}^{T}\bA_t, & \bA_t&=d_t\,(\bar\bx_t^{*}\bar\bx_t^{\T})\otimes\bI_{N_r},\nonumber\\
\bb&=\textstyle\sum_{t=1}^{T}\bb_t, & \bb_t&=d_t\,\bar\bx_t^{*}\otimes\by_t,
\label{eq:lik}
\end{align}
where $d_t=(\sigma^2+\sum_bv_{bt}e_b)^{-1}$ is the effective noise precision of column $t$ and $e_b$ the mean energy of an entry of column $b$ of $\bH$ under the previous belief.
As a function of $\bh$, this site, the Gaussian message from the pilots and the decoded data, has precision $\bG$ and mean $\bG^{-1}\bb$, a weighted least-squares estimate.
Module~H returns the prior site $(\bLambda,\bfeta)$, a factor of the same form computed from $(\bG,\bb)$ and the prior (Section~\ref{sec:problem}).
Their product is the channel belief $\mathcal{CN}(\hat\bh,\bSigma)$ with
\begin{equation}
\bSigma=(\bLambda+\bG)^{-1},\qquad \hat\bh=\bSigma(\bfeta+\bb).
\label{eq:moduleH}
\end{equation}

\textit{Detector:}
The detector treats the remaining channel error as noise, so the prior site shapes detection as well.
In data column $t$, it drops the column's own term (leave-one-out) from \eqref{eq:moduleH}, so as not to use $\by_t$ twice, but keeps the prior site: $\bSigma_{\setminus t}=(\bLambda+\bG-\bA_t)^{-1}$ and $\hat\bh_{\setminus t}=\bSigma_{\setminus t}(\bfeta+\bb-\bb_t)$.
With the decoder's extrinsic $\bx_t\sim\mathcal{CN}(\br_t,\bD_t)$ as the symbol prior (initially $\br_t=\mathbf 0$ and $\bD_t=\bI$), LMMSE detection uses $\hat\bH_{\setminus t}=\mathrm{unvec}(\hat\bh_{\setminus t})$ and folds the channel error into the noise covariance
\begin{equation}
\bXi_t=\sigma^2\bI+\sum_{b,b'}\big[\br_t\br_t^{\Hr}+\bD_t\big]_{b,b'}\,\bSigma_{\setminus t}^{[b,b']},
\label{eq:xi}
\end{equation}
where $\bSigma_{\setminus t}^{[b,b']}$ is the $(b,b')$th $N_r\times N_r$ block of $\bSigma_{\setminus t}$.
Each symbol's LMMSE marginal divided by its prior is the extrinsic message $(\tilde r_i,\tilde\tau_i)$ of code symbol~$i$ after deinterleaving.

\textit{Decoder:}
The decoder returns bit decisions and two kinds of soft symbol moments, a-posteriori ones for channel estimation and extrinsic ones for detection.
Log-domain maximum a-posteriori (log-MAP) BCJR gives the a-posteriori moments $(\bar x_i,v_i)$, which enter \eqref{eq:lik} undamped in the next iteration.
The extrinsic moments $r_i=(\bar x_i-\alpha_D\tilde r_i)/(1-\alpha_D)$ and $\tau_i=\alpha_D\tilde\tau_i/(1-\alpha_D)$ use one Onsager coefficient $\alpha_D$, the mean of $v_i/\tilde\tau_i$, as in score-based vector approximate message passing (SC-VAMP)~\cite{wadayama2026score,wadayama2026three}.
Interleaved and damped with weight 0.7 on the new value, they give the mean $\br_t$ and the diagonal covariance $\bD_t$ of the symbols in column~$t$.

\textit{Perfect channel state information (CSI):}
As a reference, not a bound, \pcsi{} runs the same detector and loop given the true channel ($\hat\bH_{\setminus t}=\bH$, $\bXi_t=\sigma^2\bI$).

\subsection{Problem Statement}\label{sec:problem}
The problem we address is to build module~H for a learned diffusion prior (Fig.~\ref{fig:problem}).
As in EP~\cite{minka2001expectation}, module~H should return the prior site for which the belief \eqref{eq:moduleH} has the mean and covariance of the channel posterior
\begin{equation}
p(\bh\,|\,\bG,\bb)\propto p(\bh)\exp(-\bh^{\Hr}\bG\bh+2\Re\{\bb^{\Hr}\bh\})
\label{eq:target}
\end{equation}
under the prior $p(\bh)$ and the likelihood site.
Gaussian and GMM priors give this site in closed form (Section~\ref{sec:gmm}).
A diffusion network, in contrast, returns only a mean, $\mathbb{E}[\bh\,|\,\bq]$, and only for a query $\bq=\bh+\mathbf n$ with $\mathbf n\sim\mathcal{CN}(\mathbf 0,\nu\bI)$, i.e., the isotropic $(\bG,\bb)=(\bI/\nu,\bq/\nu)$ in \eqref{eq:target}.
Our $\bG$ is isotropic only in the first iteration, $\bG=(T_p/\sigma^2)\bI$; the decoded data then make it anisotropic.
Section~\ref{sec:prior} forms the site from the network at one isotropic query per iteration.

\section{Learned Diffusion Prior as an EP Site}\label{sec:prior}
Module~H computes its site by the EP update~\cite{minka2001expectation} from the cavity, the product of the other sites, here the likelihood site $(\bG,\bb)$ alone.
It multiplies the prior by the cavity, giving \eqref{eq:target}, projects the product onto a Gaussian with the same mean and covariance, and divides that Gaussian by the cavity.

\subsection{Closed-Form Sites: Gaussian and GMM}\label{sec:gmm}
Both priors give the EP update in closed form: under the \gauss{}, \eqref{eq:moduleH} is the LMMSE posterior, and under the \gmm{}, \eqref{eq:target} is a Gaussian mixture with closed-form moments.
The \gauss{} uses the sample covariance $\hat\bC$ of the training channels: $\bLambda=\hat\bC^{-1}$ and $\bfeta=\mathbf 0$.
For a GMM $p(\bh)=\sum_{k=1}^{K}w_k\,\mathcal{CN}(\mathbf 0,\bC_k)$ with weights $w_k$, module~H computes the mean $\vm_{\rm GMM}$ and covariance $\mathbf V_{\rm GMM}$ of \eqref{eq:target} and divides $\mathcal{CN}(\vm_{\rm GMM},\mathbf V_{\rm GMM})$ by the cavity, which gives the closed-form site $\bLambda=\mathbf V_{\rm GMM}^{-1}-\bG$ and $\bfeta=\mathbf V_{\rm GMM}^{-1}\vm_{\rm GMM}-\bb$.
As $\bLambda$ can be indefinite (a difference of precisions), a site-precision floor, also used by the learned site, raises its eigenvalues below $\lambda_{\min}=10^{-6}$ to $\lambda_{\min}$ and leaves $\bfeta$ unchanged.
Unless it is active, \eqref{eq:moduleH} reproduces $\vm_{\rm GMM}$ and $\mathbf V_{\rm GMM}$.

The \gmm{}~\cite{koller2022asymptotically} is fitted by expectation maximization (EM) to the same $N'$ training channels, with full or Kronecker components $\bC_k=\bC_k^{\rm t}\otimes\bC_k^{\rm r}$ of transmit and receive factors.
Its family and $K$ are chosen by the log-likelihood on 5000 independent validation channels, never by BLER.

\subsection{Learned Diffusion Prior Site}\label{sec:site}
Module~H first condenses the cavity into a query $\bq$ that acts as one observation of $\bh$ in isotropic noise of variance $\nu_q$.
As in the LMMSE step of vector approximate message passing (VAMP)~\cite{rangan2019vector}, it multiplies the cavity by the network's message $\mathcal{CN}(\bh_E,\nu_E\bI)$ from its previous call.
It then projects the product onto an isotropic Gaussian and divides out $(\bh_E,\nu_E)$:
\begin{align}
&\bSigma_L=(\bI/\nu_E+\bG)^{-1},\quad \bh_L=\bSigma_L(\bh_E/\nu_E+\bb),\nonumber\\
&\alpha_L=\frac{\mathrm{tr}\,\bSigma_L}{N\nu_E},\quad
\bq=\frac{\bh_L-\alpha_L\bh_E}{1-\alpha_L},\quad
\nu_q=\frac{\alpha_L\,\nu_E}{1-\alpha_L}.
\label{eq:query}
\end{align}
The projection keeps the mean $\bh_L$ and the average variance $\alpha_L\nu_E$ of the product, and clipping $\alpha_L$ to $[\delta,1-\delta]$ with $\delta=10^{-6}$ keeps \eqref{eq:query} finite.
In the first call, $\bh_E=\mathbf 0$ and $\nu_E=\mathrm{tr}\,\hat\bC/N$; later calls form $(\bh_E,\nu_E)$ as described after \eqref{eq:site}.

The network then gives the mean $\vm$ and covariance $\bSigma_H$ of $\bh$ given $\bq$.
Its output is only the Tweedie mean; the covariance comes from its Wirtinger Jacobian:
\begin{equation}
\vm=\bq+\nu_q\nabla_{\bq^*}\log p_{\nu_q}(\bq),\qquad \bJ=\partial\vm/\partial\bq^{\T},
\label{eq:tweedie}
\end{equation}
where $\nabla_{\bq^*}$ is the Wirtinger gradient and $p_\nu$ is the density of $\bh+\mathcal{CN}(\mathbf 0,\nu\bI)$, whose score the network approximates.
For an exact minimum mean-square error (MMSE) denoiser, $\bJ=\mathrm{Cov}[\bh\,|\,\bq]/\nu_q$ is Hermitian positive semidefinite (PSD).
As in~\cite{rozet2024learning}, the denoiser belief is therefore $\mathcal{CN}(\vm,\bSigma_H)$ with $\bSigma_H=\nu_q\,\mathrm{Herm}(\bJ)$, $\mathrm{Herm}(\bJ)=(\bJ+\bJ^{\Hr})/2$: a proper Gaussian that ignores the pseudo-covariance $\partial\vm/\partial\bq^{\Hr}$.

Dividing the denoiser belief $\mathcal{CN}(\vm,\bSigma_H)$ by the Gaussian $\mathcal{CN}(\bq,\nu_q\bI)$ of the query gives the matrix site
\begin{equation}
\bLambda=\bSigma_H^{-1}-\bI/\nu_q,\qquad \bfeta=\bSigma_H^{-1}\vm-\bq/\nu_q.
\label{eq:site}
\end{equation}
To the network, $\mathcal{CN}(\bq,\nu_q\bI)$ is the cavity, i.e., $(\bG,\bb)=(\bI/\nu_q,\bq/\nu_q)$ in \eqref{eq:target}, so, within proper Gaussians, \eqref{eq:site} is the exact EP site for an MMSE denoiser when no floor is active.

For its next call, module~H projects the denoiser belief onto an isotropic Gaussian and divides out the query: $\bh_E=(\vm-\alpha_H\bq)/(1-\alpha_H)$ and $\nu_E=\alpha_H\nu_q/(1-\alpha_H)$, with $\alpha_H=\mathrm{tr}\,\bSigma_H/(N\nu_q)$ clipped like $\alpha_L$.
The controls in Section~\ref{sec:source} use this $(\bh_E,\nu_E)$ as the scalar site $\bLambda=\bI/\nu_E$, $\bfeta=\bh_E/\nu_E$.

The learned site has the interface of the GMM site, $(\bG,\bb)$ in and $(\bLambda,\bfeta)$ out, but sees the cavity only through the query and is combined with the full likelihood \eqref{eq:lik} only in \eqref{eq:moduleH}.
Each comparison thus pairs a prior with its own site computation, which does not favor the \ours{}: the GMM site, computed from $(\bG,\bb)$ itself, is more accurate.

Lemma~\ref{lem:query} bounds the noise levels that the network must cover: under our pilots, the query variance never exceeds $\sigma^2/T_p$.
Its condition holds at every SNR we test, since $\mathbb{E}\|\bH\|_F^2=N$ makes the initial $\nu_E$ close to 1.
\begin{lemma}\label{lem:query}
Let $\bX_p\bX_p^{\Hr}=T_p\bI$ and $\delta/(1-\delta)<\bar cT_p/\sigma^2<(1-\delta)/\delta$, where $\bar c=\mathrm{tr}\,\hat\bC/N$ is the initial~$\nu_E$.
Then the query \eqref{eq:query} has $\nu_q\le\sigma^2/T_p$ in every outer iteration, with equality in the first, where $\bq$ is the least-squares pilot estimate.
\end{lemma}
\begin{IEEEproof}
The pilots add $(T_p/\sigma^2)\bI$ to $\bG$, and every $\bA_t\succeq\mathbf 0$, so before clipping, $\alpha_L\le(1+\nu_ET_p/\sigma^2)^{-1}$ and $\nu_q\le\sigma^2/T_p$.
Clipping at $1-\delta$ lowers $\nu_q$.
Clipping at $\delta$ gives at most the previous $\nu_q$, as $\alpha_H\le1-\delta$.
In the first iteration, $\bG=(T_p/\sigma^2)\bI$ and $\bh_E=\mathbf 0$, so $\alpha_L$ is not clipped and $\bq=\bG^{-1}\bb$.
\end{IEEEproof}

In the first iteration, the channel belief is the denoiser belief itself: by Lemma~\ref{lem:query}, $\bG=\bI/\nu_q$ and $\bb=\bq/\nu_q$, so \eqref{eq:moduleH} with \eqref{eq:site} returns $\hat\bh=\vm$ and $\bSigma=\bSigma_H$ unless the site-precision floor is active.
The loop thus starts from the pilot-only use of the prior (Section~\ref{sec:source}) and refines it with the decoded data.

Like the \gmm{}, the network, a diffusion transformer (DiT)~\cite{peebles2023scalable}, learns from the $N'$ training channels alone.
It is trained by $\epsilon$-prediction denoising score matching~\cite{vincent2011connection} to predict the normalized noise $\hat{\boldsymbol\epsilon}\approx(\bq-\bh)/\varsigma$ at level $\varsigma=\sqrt{\nu/2}$, the noise standard deviation per real dimension, so that $\vm=\bq-\varsigma\hat{\boldsymbol\epsilon}$ in \eqref{eq:tweedie}.
The receiver then uses an exponential moving average (EMA) of the trained weights.

Training noise levels are those that the loop queries: $\log\varsigma$ is uniform between the 1st and 99th percentiles of the query levels on the first 64 test blocks per SNR.
They come from running the loop with the LMMSE denoiser of a Gaussian prior in place of the network, without recording BLER.

\subsection{Indefinite Jacobians and the Hermitian-PSD Floor}\label{sec:psd}
If $\mathrm{Herm}(\bJ)$ has a negative eigenvalue, the site-precision floor cannot keep the belief \eqref{eq:moduleH} bounded.
A learned Jacobian need not be PSD~\cite{rozet2024learning,hen2026jacobian}.
With the eigenpairs $(\lambda_i,\bu_i)$ of $\mathrm{Herm}(\bJ)$, \eqref{eq:site} gives the eigenvalue $\Lambda_i$ of $\bLambda$ along $\bu_i$ and the component $\bu_i^{\Hr}\bfeta$,
\begin{equation}
\Lambda_i=\frac{1-\lambda_i}{\nu_q\lambda_i},\qquad
\bu_i^{\Hr}\bfeta=\frac{\bu_i^{\Hr}\vm}{\nu_q\lambda_i}-\frac{\bu_i^{\Hr}\bq}{\nu_q}.
\label{eq:eig}
\end{equation}
For $\lambda_i<0$, the site-precision floor raises $\Lambda_i$ to $\lambda_{\min}$, but the first term of $\bu_i^{\Hr}\bfeta$ grows without bound as $\lambda_i\to0^-$; as $\bLambda+\bG$ in \eqref{eq:moduleH} stays bounded, the belief mean can diverge along $\bu_i$.

Flooring $\mathrm{Herm}(\bJ)$ itself keeps both quantities in \eqref{eq:eig} bounded.
The \ours{} therefore applies the Hermitian-PSD floor: it projects $\mathrm{Herm}(\bJ)$ in $\bSigma_H$ onto the Hermitian matrices with eigenvalues at least the same $\lambda_{\min}$~\cite{higham1988computing},
\begin{equation}
\bU\,\mathrm{diag}(\boldsymbol\lambda)\bU^{\Hr}\ \mapsto\ \bU\,\mathrm{diag}\bigl(\max(\boldsymbol\lambda,\lambda_{\min})\bigr)\bU^{\Hr}.
\label{eq:psd}
\end{equation}
Along a floored direction, $\Lambda_i^{-1}\bu_i^{\Hr}\bfeta\approx\bu_i^{\Hr}\vm$, so the belief stays at the denoiser output.

The Hermitian-PSD floor is necessary in practice: on Sparse specular $8\times4$ at the default $N'$, the \unrep{}, the same network without \eqref{eq:psd}, has BLER 0.593 at $-3$\,dB.
Even the \gauss{} fails less than it at $-3$, 0, and 3\,dB (Section~\ref{sec:setup}), whereas with \eqref{eq:psd} the network has BLER 0.145 at $-3$\,dB.

\section{Simulation Results}\label{sec:results}
\begin{figure*}[t]
\centering
\includegraphics[scale=0.85]{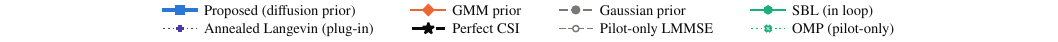}\\[1pt]
\makebox[0.84\textwidth]{%
\subfloat[]{\includegraphics[scale=0.8]{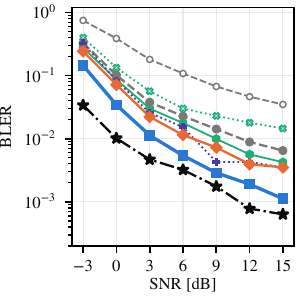}\label{fig:headline-a}}\hfill
\subfloat[]{\includegraphics[scale=0.8]{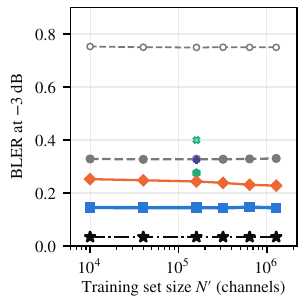}\label{fig:headline-b}}\hfill
\subfloat[]{\includegraphics[scale=0.8]{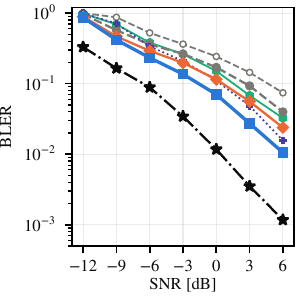}\label{fig:headline-c}}}
\caption{BLER after 16 outer iterations (one for Pilot-only LMMSE) on (a), (b)~Sparse specular $8\times4$ and (c)~3GPP UMi 28\,GHz $32\times4$: (a), (c)~versus SNR at $N'=1.6\times10^5$; (b)~at $-3$\,dB versus $N'$, with last-epoch EMA weights at every $N'$. From 6\,dB on, (a) uses 20\,480 new blocks per SNR to resolve lower BLERs (2560 for annealed Langevin estimation); (c) uses 2560 shared blocks throughout. Of the 16 baselines of Section~\ref{sec:setup}, (a) shows six. In (b), the training recipe passed the quality gate of Section~\ref{sec:setup} from $N'=1.6\times10^5$ on, the \gmm{} has $K=512$, 2048, and 1024 at the three smallest $N'$ and 4096 above, and the one-point curves were run at $N'=1.6\times10^5$ only. In (c), the \gmm{} has $K=8192$ (the largest fitted), and OMP was not run. SBL (in loop) in (c) is tuned on its loop BLER on development blocks, with its oversampling and update count both at the lower end of the searched range.}
\label{fig:headline}
\end{figure*}

\subsection{Setup and Evaluation Protocol}\label{sec:setup}
Every receiver decodes the same 2560 test blocks at each SNR of a 3-dB grid from $-3$ to 15\,dB ($-12$ to 6\,dB at $N_r=32$), and all comparisons use these blocks.
A block fails if any information bit is wrong after the 16 outer iterations or the receiver raises an exception.
Except for the training noise levels of Section~\ref{sec:site}, tuning uses development blocks, disjoint from the test blocks.

The network's training recipe passed a quality gate before any BLER was computed.
This gate trains the same recipe on an $8\times4$ testbed whose prior, a mixture of 1024 Gaussians, is known, and checks the errors of the resulting denoiser, score, and Jacobian against preset limits.
The recipe passed it for the Sparse specular $8\times4$ setting at $N'\ge1.6\times10^5$; smaller $N'$ and the other settings, which reuse this recipe, are measured with the same protocol.
The network holds out 10\% of its $N'$ training channels for validation, whereas the \gmm{} is fitted to all $N'$ and selected on the 5000 validation channels of Section~\ref{sec:gmm}, which favors the \gmm{}.

The \gmm{} is selected as in Section~\ref{sec:gmm} among fits with up to 4096 components (8192 in Section~\ref{sec:headline}).
Except for Sparse specular $8\times4$ at $N'=10^4$ and Sparse specular $32\times4$, the selected $K$ is the largest fitted.

Each comparison is decided by paired sign tests at three tested SNRs.
For a baseline~$X$, these are the SNRs at which the BLER of~$X$ lies in $[0.005,0.9]$ and is closest to 0.1 on a log scale.
At each, an exact two-sided sign test uses the blocks that fail with only one receiver.
One receiver \emph{fails less} than the other if $p<0.05$ in its favor at two or more of these SNRs.

The SNR gain $\Delta_{0.1}$ over~$X$ is the SNR of~$X$ minus that of the \ours{} at which the log-interpolated BLER first drops below 0.1.
Against the \gmm{}, the relative measure $\mathcal{R}=(F_{\mathrm{GMM}}-F_{\mathrm{P}})/(F_{\mathrm{GMM}}-F_{\mathrm{CSI}})$ is the fraction of the GMM-to-\pcsi{} gap that the \ours{} closes.
Here $F_{\mathrm{GMM}}$, $F_{\mathrm{P}}$, and $F_{\mathrm{CSI}}$ count the blocks on which the \gmm{}, the \ours{}, and \pcsi{} fail, summed over the three tested SNRs.
Both carry 90\% confidence intervals (CIs, in brackets) from paired bootstraps over the shared test blocks (2000 resamples); across arrays or training-set sizes the blocks differ, so those intervals are unpaired.
Results marked report-only are not part of the comparisons defined above.

We compare with 16 baselines in four groups of four, five, four, and three.
\emph{GMM:} the \gmm{} and its three variants: with a scalar site (Section~\ref{sec:source}), with $K=32$, and used on the pilots only (below).
\emph{Loop receivers:} the \gauss{}; a turbo LMMSE receiver, the \gauss{} without leave-one-out; bilinear generalized approximate message passing (BiG-AMP)~\cite{parker2013bigamp} with an i.i.d.\ channel prior; and SC-VAMP with an Onsager or log-likelihood-ratio (LLR) extrinsic.
SC-VAMP adapts the known-channel receiver of~\cite{wadayama2026three} to LMMSE channel estimation.
\emph{Pilot-only}, with the channel estimate fixed after one application while detection and decoding iterate: Pilot-only LMMSE (the \gauss{} after only one outer iteration), the \ours{}, and annealed Langevin estimation.
The last, the posterior sampling of~\cite{arvinte2023mimo} with our network, has the channel belief $\mathcal{CN}(\hat\bh,v\bI)$ with $v=10^{-8}$ (plug-in) or its mean-square error (MSE) on development blocks (error-aware).
\emph{Sparse:} sparse Bayesian learning (SBL)~\cite{tipping2001sparse} as the prior in module~H of the same loop (re-estimated every iteration) or on the pilots, and orthogonal matching pursuit (OMP)~\cite{pati1993orthogonal} on the pilots.
All three use an oversampled \mbox{2-D} DFT dictionary; its oversampling, SBL's update count, and OMP's sparsity are tuned on the pilot normalized MSE (NMSE) on development blocks.
The oversampling is at the end of the searched grid for all but SBL on 3GPP UMi 28\,GHz.
The code is at \href{https://github.com/TaeJun1999/diffusion-ep-site}{\nolinkurl{github.com/TaeJun1999/diffusion-ep-site}}.

\subsection{Gain Over the GMM Prior and Other Baselines}\label{sec:headline}
On Sparse specular $8\times4$ with $N'=1.6\times10^5$, the \ours{} fails less than the \gmm{} at all three tested SNRs, $-3$, 0, and 3\,dB (Fig.~\ref{fig:headline}(a)).
At $-3$\,dB, it lowers the BLER from 0.243 to 0.145, while \pcsi{} has 0.034, and its $\Delta_{0.1}$ is 1.41\,dB [$+1.22$, $+1.64$].
The gain holds at every tested training-set size.
At each of six sizes from $N'=10^4$ to ${1.28\times10^6}$, it fails less than the \gmm{} at all three tested SNRs (Fig.~\ref{fig:headline}(b)).
Trained on $N'=1.6\times10^5$ channels, it needs 1.11\,dB less SNR for BLER 0.1 than the \gmm{} fitted to eight times as many (unpaired CI [$+0.86$, $+1.36$]).

At $N'=1.6\times10^5$ on Sparse specular $8\times4$, the \ours{} fails less than each of the 16 baselines, including three sparse estimators.
The closest are the pilot-only use of the same prior ($\Delta_{0.1}=0.92$\,dB), the \gmm{} (1.41), and error-aware annealed Langevin estimation (1.54 [$+1.37$, $+1.74$]); for SBL (in loop), $\Delta_{0.1}$ is 1.86\,dB [$+1.64$, $+2.10$].
On 3GPP UMi 28\,GHz $8\times4$, it again fails less than all 16, with $\Delta_{0.1}=0.60$\,dB [$+0.33$, $+0.87$] over the \gmm{}.

The gain over the \gmm{} is larger at $N_r=32$ than at~8.
With $N_r=8$, 16, and 32 on Sparse specular and 3GPP UMi 28\,GHz, the \ours{} fails less than the \gmm{} in all six settings, each with its own network, \gmm{}, and tested SNRs.
Against these \gmm{}s ($K\le4096$), its $\Delta_{0.1}$ rises from 1.41 to 2.99\,dB on Sparse specular and from 0.60 to 2.29\,dB on 3GPP UMi 28\,GHz.
The fraction $\mathcal R$ of the GMM-to-\pcsi{} gap it closes, summed over each setting's tested SNRs, is 0.509, 0.823, and 0.811 on Sparse specular and 0.150, 0.287, and 0.432 on 3GPP UMi 28\,GHz.
Both increases of $\mathcal R$ from $N_r=8$ to 32 have $p<0.001$ in an unpaired bootstrap over blocks; in absolute terms, $F_{\mathrm P}-F_{\mathrm{CSI}}$ goes from 363 to 206 and from 497 to 472 (report-only).
At $32\times4$ on 3GPP UMi 28\,GHz, the advantage also holds against a larger \gmm{}, with $K=8192$ instead of 4096 (Fig.~\ref{fig:headline}(c)).
It has a higher validation likelihood but is again the largest $K$ fitted.
Against it, the \ours{} still fails less at all three tested SNRs, with $\Delta_{0.1}=2.17$\,dB [$+1.83$, $+2.50$] instead of 2.29\,dB.

\subsection{Where the Gain Comes From}\label{sec:source}
On Sparse specular and 3GPP UMi 28\,GHz $8\times4$, the turbo loop fails less than the pilot-only use of the same prior.
It needs 0.92\,dB [$+0.77$, $+1.07$] and 0.37\,dB [$+0.18$, $+0.57$] less SNR for BLER 0.1.
Pilot-only use, as in~\cite{arvinte2023mimo}, applies module~H once; in the loop it runs 16 times, so this gain combines re-estimation with the decoded data and repeated application of the prior.

The gain over the \gmm{} comes from the diffusion prior in module~H, not from the matrix site; two controls on Sparse specular $8\times4$ show this.
Each gives one prior only the scalar site of Section~\ref{sec:site} in place of its matrix site.
For the network, $\bSigma_H$ enters only through $\alpha_H$, without the floor \eqref{eq:psd}; for the \gmm{}, its closed-form moments under the isotropic query replace $\vm$ and $\bSigma_H$.
The resulting \scdiff{} still fails less than the \gmm{} with its matrix site at all three tested SNRs and attains BLER 0.1 at a 1.27\,dB [$+1.08$, $+1.50$] lower SNR.
The \gmm{} keeps its matrix site: no tested SNR gives $p<0.05$ in favor of its scalar site.
The \ours{} keeps the matrix site, so that the detector's noise covariance \eqref{eq:xi} reflects the full $\bSigma_H$.

\subsection{Cost}\label{sec:cost}
Per call, the learned site evaluates the network with $2N$ vector--Jacobian products and adds the $N\times N$ eigendecomposition of \eqref{eq:psd}, whereas the GMM site inverts $K$ matrices of size $N\times N$.
In our research implementation on one CPU thread, a block at $-3$\,dB takes 1.0\,s with the \ours{} and 0.76\,s with the \gmm{} ($K=1024$) on Sparse specular $8\times4$.
On 3GPP UMi 28\,GHz $32\times4$, it takes 85\,s, against 90 and 184\,s for $K=4096$ and 8192.

\section{Conclusion}\label{sec:conclusion}
We proposed a matrix-valued EP site that lets a coded MIMO turbo receiver use a learned diffusion prior as its channel prior.
At an isotropic query, one network gives the channel's mean and covariance, and a Hermitian-PSD floor bounds the site even when the network's Jacobian is indefinite.
On a controlled sparse specular model and a 3GPP model, the \ours{} failed less than a \gmm{} learned from the same training channels and, at $8\times4$, than each of 16 baselines.
Future work includes cutting the cost of the Jacobian for large arrays and extending the site to links with low-density parity-check (LDPC) codes over frequency-selective channels.

\bibliographystyle{IEEEtran}
\bibliography{references}

\end{document}

%% file: figures/F1_receiver_eqnums.tex
\figeqcheck{eq:lik}{4}\figeqcheck{eq:xi}{6}\figeqcheck{eq:query}{8}\figeqcheck{eq:site}{10}\figeqcheck{eq:psd}{12}